# Ionic Clusters vs. Shear Viscosity in Aqueous Amino Acid Ionic Liquids

Vitaly V. Chaban and Eudes Eterno Fileti

Instituto de Ciência e Tecnologia, Universidade Federal de São Paulo, 12247-014, São José dos Campos, SP, Brazil

**Abstract**. Aqueous solutions of amino acid ionic liquids (AAILs) are of high importance for applications in protein synthesis and solubilization, enzymatic reactions, templates for synthetic study, etc. This work employs molecular dynamics simulations using our own force field to investigate shear viscosity and cluster compositions of three 1-ethyl-3-methylimidazolium (emim) amino acid salts: [emim][ala], [emim][met], and [emim][trp] solutions (2, 5, 10, 20, 30 mol%) in water at 310 K. We, for the first time, establish simple correlations between cluster composition, on one side, and viscosity, on another side. We argue that knowledge about any of these properties alone is enough to derive insights regarding the missed properties, using the reported correlations. The numerical observations and qualitative correlations are discussed in the context of chemical structure of the amino acid anions, [ala], [met], and [trp]. The reported results will enhance progress in the efficient design and applications of AAILs and their solutions.

**Introduction**

New organic and inorganic ions are being currently introduced giving rise to multiple families of ionic liquids (ILs).[1-4] ILs provide an exciting platform for physical chemical engineering stimulating a variety of applications.[5-14] Amino acid based ionic liquids (AAILs) exhibit a huge potential in the field of protein chemistry, enzymatic reactions, and templates for synthetic study.[15-24] Both cation and anion of a single ionic liquid can originate from biological species. Due to compatibility with living cells, such liquids are referenced to as "bio-ionic".[25] They may eliminate toxicity hazards, which characterize many conventional ionic liquids.[10] The strong hydrogen bonding ability of AAILs, as compared to other ionic liquids, makes them more demanding in synthetic, pharmaceutical, and medicinal chemistry.[14, 26-28] AAILs can be used as templates during peptide synthesis, which constitutes a separate vivid research field per se. Additionally, AAILs are interesting as chiral solvents and reactants to dissolve and stabilize biomolecules, such as cellulose, nucleic acids, carbohydrates and other species of primary biological importance.[29] The presence of certain amounts of water is implied in all these applications, as water is omnipresent in living systems. The potential use of AAILs can only be fully realized when the detailed atomistic-precision information about physical chemical properties, structures and interactions in their condensed phase becomes available.[4, 16, 30-33] Apart from the microscopic structures of pure AAILs, behavior of these liquids in aqueous medium is of undoubted interest. In particular, the extent of miscibility of AAILs and water are important for many practical applications. We consider AAILs based on 1-ethyl-3-methylimidazolium cation (emim).

The present work reports shear viscosity and ionic clusters in the 2, 5, 10, 20, and 30 mol% aqueous solutions of [emim][ala], [emim][met], and [emim][trp] at 310 K. Molecular dynamics (MD) simulations based on pairwise empirical potentials are used to derive physical chemical properties from atomistic trajectories. We show that percentage of ion pairs and maximum cluster size in the MD system are connected to viscosity via simple mathematical expressions.

The established correlations allow to efficiently predict transport properties based on structure data, and contrariwise.

**Simulation Details**

Classical molecular dynamics (MD) simulations using conventional additive interaction potentials were performed employing our own force field (FF) for AAILs. TIP3P model of water was used to provide internal compatibility.[34] See Ref. [35] for details of the FF derivation and compatibility considerations. 1-ethyl-3-methylimidazolium alanine, [emim][ala], 1-ethyl-3-methylimidazolium methionine, [emim][met], 1-ethyl-3-methylimidazolium tryptophan, [emim][trp], were dissolved in water to produce 2, 5, 10, 20, and 30 mol% aqueous solutions (Figure 1). These particular AAILs were selected due to possessing significantly different amino acid residues. Indeed, alanine is a small molecule, which is largely hydrophilic. We expect perfect miscibility of [emim][ala] with water. In turn, methionine is more hydrophobic, while possessing sulfur. The sulfur atom is spatially separated from the hydrophilic moiety by the two methylene groups. Tryptophan is a relatively large molecule, which possesses two fused aromatic rings. Although five-membered ring contains a nitrogen atom, it does not alter a hydrophobic nature of the entire radical. All introduced anions were obtained by deprotonating carboxyl groups of the respective amino acid. The list of the simulated systems is provided in Table 1.

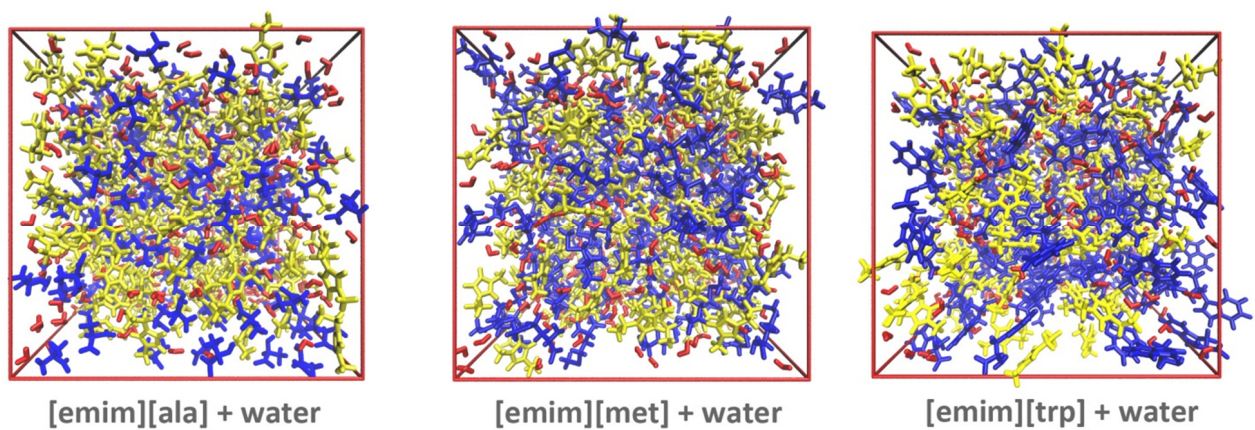

Figure 1. Computational unit cells for the three selected amino acid ionic liquids (20 mol%). 1-ethyl-3-methylimidazolium cations are yellow, amino acid anions are blue, and water molecules are red.

Table 1: MD systems considered in the present work. The quantity of AAIL ion pairs per system was selected with respect to cation and anion sizes. The number of water molecules was selected to obtain the required molar concentration of each solution

| # | AAIL | # ion pairs | # water molecules | # interaction sites | mol% (AAIL) |
|---|------|-------------|-------------------|---------------------|-------------|
| 1 |            | 30  | 1470 | 5,340 | 2  |
| 2 |            | 60  | 1140 | 5,280 | 5  |
| 3 | [emim][ala] | 100 | 900  | 5,800 | 10 |
| 4 |            | 125 | 500  | 5,375 | 20 |
| 5 |            | 150 | 350  | 5,700 | 30 |
| 6 |            | 30  | 1470 | 5,550 | 2  |
| 7 |            | 60  | 1140 | 5,700 | 5  |
| 8 | [emim][met] | 100 | 900  | 6,500 | 10 |
| 9 |            | 125 | 500  | 6,250 | 20 |
| 10 |           | 150 | 350  | 6,750 | 30 |
| 11 |           | 30  | 1470 | 5,760 | 2  |
| 12 |           | 60  | 1140 | 6,120 | 5  |
| 13 | [emim][trp] | 75  | 675  | 5,400 | 10 |
| 14 |           | 100 | 400  | 5,700 | 20 |
| 15 |           | 120 | 280  | 6,240 | 30 |

The Cartesian coordinates were saved every 1.0 ps for future processing in accordance with relationships of statistical physics. More frequent saving of trajectory components was preliminarily tested, but no systematic accuracy improvement was found. The first 10 ns of the simulation were regarded as equilibration. The subsequent 100 ns of the simulation were used for structure analysis. All systems were simulated in the constant-pressure constant-temperature ensemble, which allows attaining a natural mass density of each model system. The equations of motion were propagated with a time-step of 2.0 fs. Such a relatively large time-step was possible due to constraints imposed on the carbon-hydrogen covalent bonds (instead of a harmonic potential). Note, the time-step of 2.0 fs is more than ten times smaller than the period of the fastest harmonic bond oscillation in any of these systems.

The electrostatic interactions were simulated using direct Coulomb law up to 1.2 nm of separation between each pair of interaction sites. The electrostatic interactions beyond 1.2 nm were accounted for by computationally efficient Particle-Mesh-Ewald (PME) method.[36] It is important to use the PME method in the case of ionic systems, since electrostatic energy beyond the cut-off usually contributes 40-60% of the total electrostatic energy. The Lennard-Jones-12-6 interactions were smoothly brought down to zero from 1.1 to 1.2 nm using the classical shifted force technique.[37] The constant temperature, 310 K, was maintained by the Bussi-Donadio-Parrinello velocity rescaling thermostat (with a time constant of 0.1 ps), which provides a correct velocity distribution for a statistical mechanical ensemble.[38] The constant pressure was maintained by Parrinello-Rahman barostat with a time constant of 1.0 ps and a compressibility constant of $4.5 \times 10^{-5}$ bar$^{-1}$.[39] The compressibility constant only determines a genuine rigidity of the barostat. It should not be numerically equal to physical compressibility of the substance to obtain a correct system density.

All molecular dynamics trajectories were propagated using the GROMACS simulation engine.[37] Analysis of structure properties was performed using the supplementary utilities distributed with the GROMACS simulation suite, where applicable. The AAILs and water were placed in cubic periodic MD boxes (Figure 1), whose initial densities were calculated to approximately correspond to ambient pressure at 310 K.

Shear viscosity of each of 15 systems was calculated from non-equilibrium MD simulations with cosine-type acceleration.[40] The subsequent energy dissipation is used to assess system viscosity. The mathematical foundations of this method are provided elsewhere.[40, 41]

Ionic clusters are defined using the single-linkage approach with a cut-off distance of 0.3 nm.[37] The coordination center of cation was chosen to be most electron deficient imidazole hydrogen atom. The coordination center of anion was chosen to be deprotonated carboxyl group. These selections were inspired by electronic structure calculations of the AAIL ion pairs in vacuum.[35] Any selected coordination center, located closer than 0.3 nm from the existing cluster, is considered to belong to the cluster.

**Results and Discussion**

Ionic structure of the AAIL solutions is characterized by means of ionic clusters formed by cations and anions together. An ion belongs to a given cluster when it exhibits at least one direct contact with any of the ions of this cluster. We select the two major descriptors: (1) the percentage of ions, which belong to ion pairs (as opposed to single ions and larger ionic clusters) and (2) the average size of the largest ionic cluster in the system. Note that the largest ionic cluster must be significantly smaller than the total number of ions in the system. Otherwise, AAIL engenders a separate phase (solid or liquid one) within water indicating limited solubility of ions. Figure 2 demonstrates the percentage of the cation-anion pairs and the size of the largest ionic cluster as a function of AAIL molar fraction. The number of ion pairs systematically decreases as more AAIL ions are supplied. Larger numbers of ions give rise to larger ionic clusters (Figure 2, bottom). In the AAIL-poor systems, 2-5 mol%, no ionic cluster contains more than 10 ions. Dependence of ionic structure on the nature of amino acid anion is clearly observed. The largest number of ion pairs exist in [emim][trp], whereas the largest ionic clusters are found in [emim][ala]. The reason for this observation probably lies in the binding strength between the cation and the anion in these AAILs and hydration of the anion (since the cation is common for all these AAILs). [ala]$^-$ is a small hydrophilic anion. Therefore, it gets hydrated and participates in a smaller percentage of ion pairs, unlike in the case of [trp]$^-$. However, [emim][ala] forms the largest ionic clusters in the AAIL-rich solutions. [met]$^-$ and [trp]$^-$, which exhibit more complicated chemical structures and, consequently, more complicated non-bonded interactions, prefer to maintain a larger number of smaller clusters.

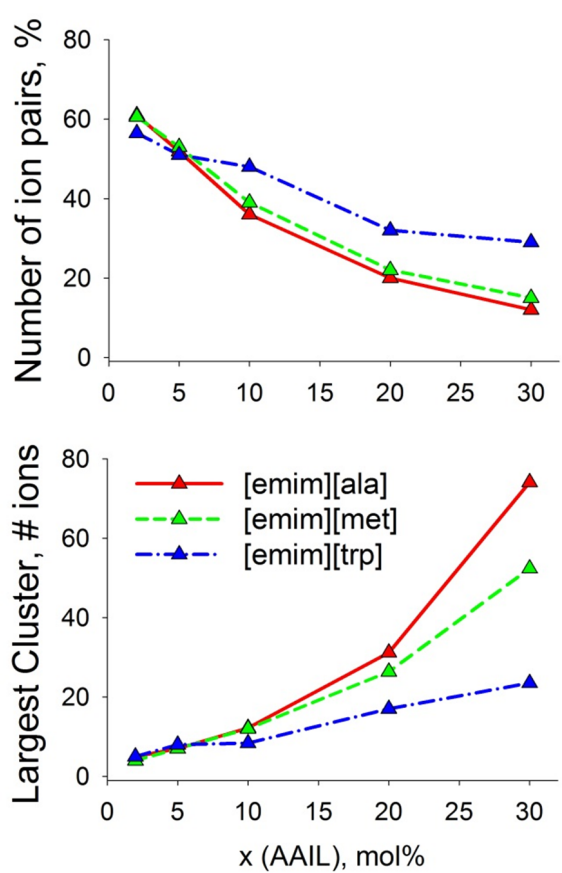

Figure 2. Number of ion pairs (top) and the size of the largest ionic cluster (bottom) vs. molar fraction in the aqueous solutions of amino acid based ionic liquids, [emim][ala], [emim][met], and [emim][trp]. See legends for system designations.

Viscosity of a liquid can be decreased by an addition of a less viscous liquid provided that these liquids are miscible over a composition of interest. Figure 3 provides shear viscosity of the aqueous solutions of the selected AAILs as a function of AAIL molar fraction. The viscosities of all AAILs exponentially increase, as the content of AAIL is linearly increased. This feature indicates good miscibility of all three AAILs and water molecules. The viscosity alterations would have been stepwise in the case of limited miscibility and formation of separate phases. The absolute values of viscosity are in concordance with our expectations based on the sizes of the anions. Provided that interactions between the cation and the anion are roughly similar in all AAILs, viscosity increases in the range [emim][ala] < [emim][met] <[emim][trp], as the size of the anion increases. Information regarding viscosity is of principal important for practical applications of these systems. Too viscous systems may complicate manipulations of them.

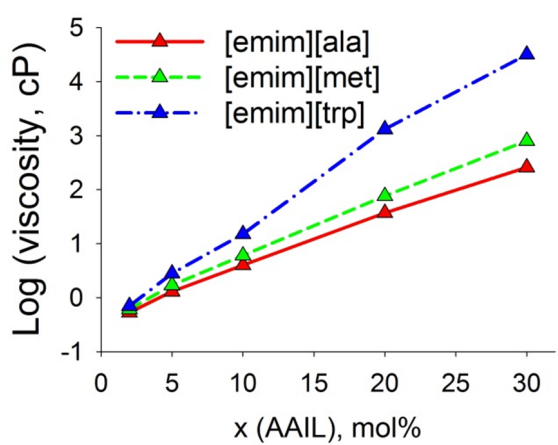

Figure 3. Shear viscosity vs. content of AAIL in the aqueous solutions of [emim][ala], [emim][met], and [emim][trp]. See legends for system designations.

Below, we establish numerical correlations between the four descriptors of aqueous solutions of [emim][ala], [emim][met], and [emim][trp]: (1) molar fraction of AAIL; (2) shear viscosity of AAIL solution; (3) percentage of ion pairs in AAIL solution; (4) the largest cluster size in AAIL solution. We demonstrate that ionic cluster structure can be in a simple way linked to shear viscosity. Figures 4-5 and Table 2 describe the fitting procedures and the corresponding parameters.

The percentage of ion pairs decreases as the first-order polynomial with respect to the logarithm of shear viscosity (Figure 4). In turn, viscosity increases exponentially with the molar content of AAIL. Table 2 shows that in the limit of small viscosity ($A_0$ parameter), the expected percentage of ion pairs in all AAILs is similar, 52.48-54.71%. The correlation coefficient, $R^2$, is reasonably high (Table 2) for this number of plotted data amounting to 0.96-0.97. The slope parameter ($A_1$) is negative (i.e. higher viscosity means less ion pairs) and increases as the size of the anion increases.

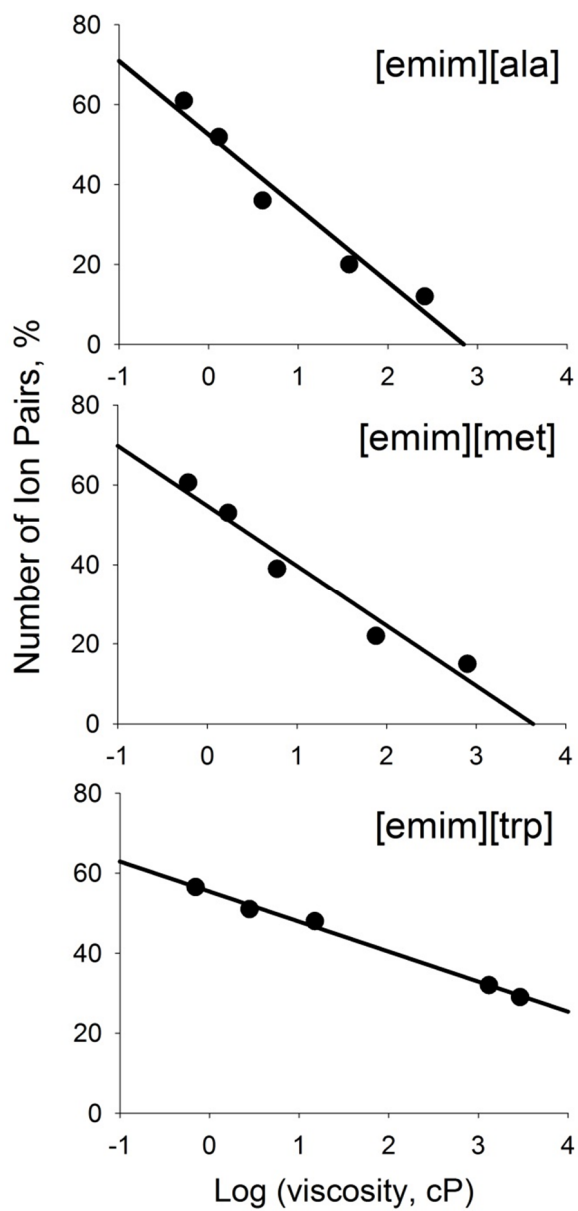

Figure 4. The established correlation between shear viscosity and the percentage of ion pairs in the aqueous solutions of AAILs, [emim][ala], [emim][met], and [emim][trp].

Table 2. Fitting equations and their parameters to correlate shear viscosity, percentage of ion pairs, and the largest cluster size in the 2, 5, 10, 20, and 30 mol% aqueous solutions of [emim][ala], [emim][met], and [emim][trp]

| AAIL | Proposed functional form and parameters | | |
|---|---|---|---|
| Percentage of ion pairs vs. log (viscosity): $A_0+A_1x$ | | | |
| | $A_0$ | $A_1$ | $R^2$ |
| [emim][ala] | 52.48 | -18.44 | 0.96 |
| [emim][met] | 54.71 | -15.06 | 0.96 |
| [emim][trp] | 54.70 | -6.150 | 0.97 |
| Size of the largest cluster vs. log (viscosity): $A_0+A_1x+A_2x^2$ | | | |
| | $A_0$ | $A_1$ | $A_2$ | $R^2$ |
| [emim][ala] | 5.72 | 0.06 | 11.5 | 0.99 |
| [emim][met] | 5.36 | 4.05 | 4.14 | 1.00 |
| [emim][trp] | 5.47 | 1.40 | 1.96 | 0.99 |

The largest cluster size is perfectly described by the second-order polynomial with respect to the logarithm of shear viscosity (Figure 5). Note excellent correlation coefficients, $R^2$, in the cases of [emim][ala], [emim][met], and [emim][trp] (Table 4). The established correlations are possible due to significant solubility of AAILs in water, whereas limited solubility would result in a biphasic system with different functional dependences. The ability to link structure and dynamic properties in a condensed matter system is important to improve predictability of those properties. Prediction of structure normally requires less extensive sampling of the phase trajectory than prediction of transport properties, especially those based on collective correlation functions, such as shear viscosity, ionic conductivity, heat conductivity. Prevalence of ion pairs in the ionic solutions (as opposed to larger ionic clusters) favors smaller viscosity, since it means better dispersion of ions throughout the system and, therefore, faster energy dissipation thanks to thermal motion.

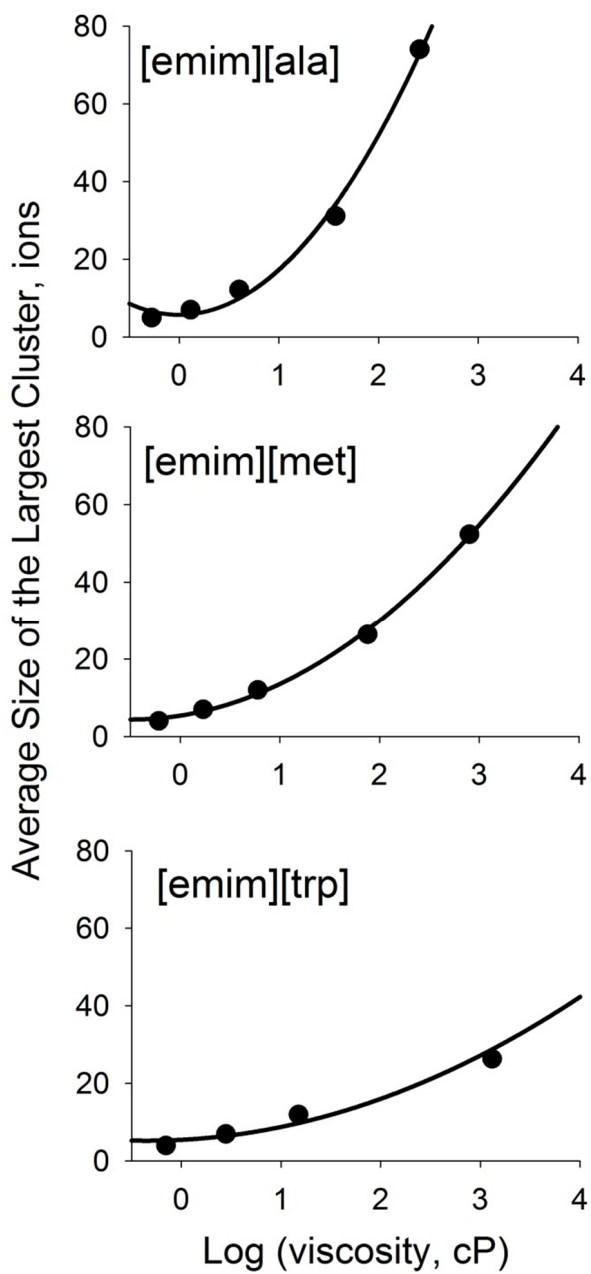

Figure 5. The established correlation between shear viscosity and the size of the largest ionic cluster in the aqueous solutions of AAILs, [emim][ala], [emim][met], and [emim][trp].

**CONCLUSIONS**

We establish a simple correlation between shear viscosity and ionic clusters in the aqueous solutions of [emim][ala], [emim][met], and [emim][trp] AAILs. In particular, we prove that viscosity of the system exponentially increases as the percentage of ion pairs linearly decreases. Similarly, the average size of the largest ionic cluster exhibits a second-order polynomial dependence on the viscosity logarithm. Viscosity itself increases exponentially, while the molar content of AAILs increases linearly. Viscosity increases faster for AAILs featuring larger amino acid anions, i.e. [tpr] > [met] > [ala]. This feature is important for understanding when aqueous solutions of AAILs are designed. Exponential viscosity decrease of more viscous liquid upon an addition of less viscous liquid is observed only when these two liquids exhibit miscibility over an entire concentration range. Note, however, that the largest cluster size increases with the content of AAIL.

The reported results are important to knowingly design novel physical chemical systems containing ionic liquids to meet practical needs. The established correlations between structure (ionic clusters) and dynamical (shear viscosity) properties can be used to estimate any of them when the other is known. Furthermore, a clear mathematical relationship permits to avoid experimental measurements at all compositions, but perform them only in a few boundary points.

**ACKNOWLEDGMENTS**

V.V.C. acknowledges research grant from CAPES (Coordenação de Aperfeiçoamento de Pessoal de Nível Superior, Brasil) under "Science Without Borders" program. E.E.F. thanks Brazilian agencies FAPESP and CNPq for support.

**CONTACT INFORMATION**

E-mail addresses for correspondence: vvchaban@gmail.com (V.V.C.); fileti@gmail.com (E.E.F.)

**TOC Graphic**

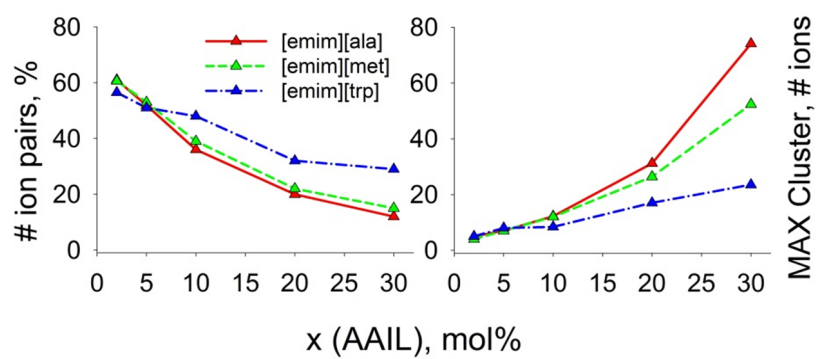